\documentclass[11pt,superscriptaddress,aps,prd,preprint]{revtex4}
\usepackage{amsmath}

\makeatletter
\usepackage[T1]{fontenc}
\usepackage{amsmath}
\usepackage{amssymb}
\usepackage{graphicx}

\usepackage{slashed}

\newcommand{\bea}{\begin{eqnarray}}
\newcommand{\eea}{\end{eqnarray}}

\begin{document}

\title{Gravitational Casimir effect at finite temperature}

\author{A. F. Santos}\email[]{alesandroferreira@fisica.ufmt.br}
\affiliation{Instituto de F\'{\i}sica, Universidade Federal de Mato Grosso,\\
78060-900, Cuiab\'{a}, Mato Grosso, Brazil}
\affiliation{Department of Physics and Astronomy, University of Victoria,\\
3800 Finnerty Road Victoria, BC, Canada}

\author{Faqir C. Khanna\footnote{Professor Emeritus - Physics Department, Theoretical Physics Institute, University of Alberta\\
Edmonton, Alberta, Canada}}\email[]{khannaf@uvic.ca}
\affiliation{Department of Physics and Astronomy, University of Victoria,\\
3800 Finnerty Road Victoria, BC, Canada}

\begin{abstract}
The energy-momentum tensor for the gravitoelectromagnetism (GEM) theory in the real-time finite temperature field theory formalism is presented. Expressions for the Casimir energy and pressure at zero and finite temperature are obtained. An analysis of the Casimir effect for the GEM field is developed.
\end{abstract}

\maketitle

\section{Introduction}

Casimir \cite{Casimir} predicted that conducting parallel plates attract due to vacuum fluctuations of the electromagnetic field. The attraction between plates is the result of electromagnetic modes due to boundary conditions or topological effects \cite{Milton, Milonni, Plunien, Bordag}. However this effect is more general and it can be defined as the force per unit area of the bounding surfaces when a quantum field is confined. 

The Casimir effect was confirmed experimentally first by Sparnaay \citep{Sparnaay} and subsequent experiments have established this effect to a high degree of accuracy \cite{Lamoreaux, Mohideen}. Recently some practical applications of this effect have been found in micro- and nanotechnologies \cite{Bordag, Lamoreaux1} involving superconductors at high temperature \citep{Bordag1, Orlando}. The finite temperature corrections to this effect are significant for plates separated by a distance of the order of micrometers \cite{Mehra}. The role of finite temperature for Casimir effect has been considered using the Thermo Field Dynamics (TFD) formalism in electromagnetic \cite{Khanna0} and Kalb-Ramond \cite{Ademir} fields.

TFD, a real-time finite temperature formalism, is a thermal quantum field theory \cite{Umezawa1, Umezawa2, Umezawa22, Khanna1, Khanna2}. The main idea is to interpret the statistical average of an arbitrary operator $A$, as the expectation value in a thermal vacuum, i.e., $\langle A \rangle=\langle 0(\beta)| A|0(\beta) \rangle$. The thermal vacuum, $|0(\beta) \rangle$, describes the thermal equilibrium of the system, where $\beta=\frac{1}{k_BT}$, and $T$ is the temperature and $k_B$ is the Boltzmann constant (we use $k_B=1$ and $\hbar=c=1$). Two elements are necessary to construct this thermal state: (i) doubling of the original Hilbert space and (ii) Bogoliubov transformations. This doubling is defined by dual conjugation rules of physical quantities represented by nontilde and tilde ($^\thicksim$) variables. The Bogoliubov transformation introduces a rotation in the tilde and nontilde variables. Here we use the TFD formalism to calculate the gravitational Casimir effect at zero and finite temperature.

The Casimir effect appears for any quantum field. Then there is a natural question: Is this phenomenon due to the gravitational field? If gravity has a quantum nature, this effect is expected for gravitational waves. For plates made of conventional materials the gravitational Casimir effect is very tiny. Although small, the Casimir effect in different gravitational models has been investigated \cite{Fulling, Parashar, Zhuk, Bezerra1, Bezerra2, Bezerra3, Bezerra4, Bezerra5}. Some of these calculations investigate whether this effect obeys the equivalence principle. Other analyses involve quantum fields (scalar and electromagnetic fields) propagating in certain gravitational backgrounds. Recently \cite{Quach} a different attempt has been proposed where plates are made of superconducting material. Using this idea and the formulation of the gravitoelectromagnetism (GEM), the gravitational Casimir effect is analyzed. In this work the Casimir energy and pressure for the GEM field are calculated and our results are compared with the electromagnetic case. 
 
GEM, a theory of gravity, is based on an analogy with electromagnetism \cite{Maxwell,Heaviside1, Heaviside2}. This idea has relevance since several experiments have been carried out to detect the gravitomagnetic contribution \cite{Wheeler, Nordtvedt, Soffel, Everitt}. There are two possible formulations for GEM: (i) based on the Weyl tensor ($C_{ijkl}$) that divides into electric and magnetic components and (ii) based on the similarity between the Maxwell and linearized Einstein equations which is valid in the framework of the weak field approximation. Here the first formulation using the Weyl tensor is considered. In this case the field equations for components of the Weyl tensor, ${\cal E}_{ij}=-C_{0i0j}$ (gravitoelectric field) and ${\cal B}_{ij}=\frac{1}{2}\epsilon_{ikl}C^{kl}_{0j}$ (gravitomagnetic field), have a structure similar to Maxwell equations.

A lagrangian formulation for GEM with gauge invariant field equations and interactions between fermions, photons and gravitons has been studied \citep{Khanna}. The interaction of the graviton with fermions and photons leads to a definition of the S-matrix, that helps us to formulate a perturbation series in order to define transition amplitudes. This leads to a symmetric energy-momentum tensor, that is used to calculate the Casimir effect for GEM field at zero and finite temperature using TFD.

This paper is organized as follows. In section II, TFD and some characteristics of this formalism are presented. In section III, a lagrangian formulation for GEM is introduced. The Stefan-Boltzmann law is derived. In section IV, the Casimir effect at zero and finite temperature is obtained. In section V, some concluding remarks are presented.

\section{Thermo Field Dynamics}

Here brief ideas of real-time finite temperature field theory (TFD) \cite{Umezawa1, Umezawa2, Santana1, Santana2} are presented. In this formalism the original Fock space ${\cal S}$ of the system is doubled leading to an expanded space ${\cal S}_T={\cal S}\otimes \tilde{\cal S}$, applicable to systems in  a thermal equilibrium state. This doubling is defined by a mapping ($^\thicksim$): ${\cal S}\rightarrow {\tilde{\cal S}}$ associating each operator say $a$, in ${\cal S}$ to two operators in ${\cal S}_T$. For an arbitrary bosonic operator ${\cal X}$ the standard doublet notation \cite{Umezawa22} is
\bea
{\cal X}^a=\left( \begin{array}{cc} {\cal X}\\ 
-\tilde{{\cal X}}^\dagger \end{array} \right).
\eea
The physical variables are described by nontilde operators.

Thermal variables are introduced by a Bogoliubov transformation, ${\cal U}(\alpha)$, that corresponds to a rotation in the tilde and non-tilde variables, such that thermal effects emerge. The Bogoliubov transformation is defined as
\bea
{\cal U}(\alpha)=\left( \begin{array}{cc} u(\alpha) & -v(\alpha) \\ 
-v(\alpha) & u(\alpha) \end{array} \right),
\eea
where $u^2(\alpha)-v^2(\alpha)=1$. These quantities $u(\alpha)$ and $v(\alpha)$ are related to the Bose distribution. The parameter $\alpha$ is associated with temperature, but, in general, it may be associated with other physical quantities.

As an example, consider a free scalar field in Minkowski space such that $diag(g^{\mu\nu})=(+1,-1,-1,-1)$. Using the Bogoliubov transformation the $\alpha$-dependent scalar field is given by
\bea
\phi(x;\alpha)&=&{\cal U}(\alpha)\phi(x){\cal U}^{-1}(\alpha),\nonumber\\
\tilde{\phi}(x;\alpha)&=&{\cal U}(\alpha)\tilde{\phi}(x){\cal U}^{-1}(\alpha).
\eea
The propagator for the $\alpha$ scalar field is
\bea
G_0^{(ab)}(x-x';\alpha)=i\langle 0,\tilde{0}| \tau[\phi^a(x;\alpha)\phi^b(x';\alpha)]| 0,\tilde{0}\rangle,
\eea
where $a, b=1,2$ and $\tau$ is the time ordering operator. Using $|0(\alpha)\rangle={\cal U}(\alpha)|0,\tilde{0}\rangle$
\bea
G_0^{(ab)}(x-x';\alpha)&=&i\langle 0(\alpha)| \tau[\phi^a(x)\phi^b(x')]| 0(\alpha)\rangle,\nonumber\\
&=&i\int \frac{d^4k}{(2\pi)^4}e^{-ik(x-x')}G_0^{(ab)}(k;\alpha),
\eea
where 
\bea
G_0^{(ab)}(k;\alpha)={\cal U}^{-1}(k;\alpha)G_0^{(ab)}(k){\cal U}(k;\alpha),
\eea
with
\bea
{\cal U}(k;\alpha)=\left( \begin{array}{cc} u(k;\alpha) & -v(;,\alpha) \\ 
-v(k;\alpha) & u(k;\alpha) \end{array} \right), \quad\quad\quad\quad G_0^{(ab)}(k)=\left( \begin{array}{cc} G_0(k) & 0 \\ 
0 & -G^*_0(k) \end{array} \right),
\eea
and 
\bea
G_0(k)=\frac{1}{k^2-m^2+i\epsilon}.
\eea
The physical information is given by $G_0^{(11)}(k;\alpha)$. Here $G_0^{(11)}(k;\alpha)\equiv G_0(k;\alpha)$, then
\bea
G_0(k;\alpha)=G_0(k)+v^2(k;\alpha)[G_0(k)-G^*_0(k)],
\eea
such that 
\bea
[G_0(k)-G^*_0(k)]=2\pi i\delta(k^2-m^2).
\eea

For the massless scalar field 
\bea
G_0(x-x')=-\frac{i}{(2\pi)^2}\frac{1}{(x-x')^2-i\epsilon}.\label{G0}
\eea

The TFD propagator, in the Heisenberg picture, is written as
\bea
G_0(x-x';\alpha=\beta)&=&i\langle 0(\beta)| \tau[\phi(x)\phi(x')]| 0(\beta)\rangle,\nonumber\\
&=&i\,\mathrm{Tr} \left\{\rho(\beta)\tau[\phi(x)\phi(x')]\right\},\nonumber\\
&=&G_0(x-x'-i\beta n_0;\beta),
\eea 
where $\rho(\beta)$ is the equilibrium density matrix for the grand-canonical ensemble and $n_0=(1,0,0,0)$. This result shows that the propagator is a periodic function with a period $\beta$ in the imaginary-time axis.

\section{GEM energy-momentum tensor}

We begin with the lagrangian formulation of GEM which is based on Maxwell-like equations \cite{Khanna}
\bea
\partial^i{\cal E}^{ij}&=&-4\pi G\rho^j,\label{01}\\
\partial^i{\cal B}^{ij}&=&0,\label{02}\\
\epsilon^{\langle ikl}\partial^k{\cal B}^{lj\rangle}+\frac{1}{c}\frac{\partial{\cal E}^{ij}}{\partial t}&=&-\frac{4\pi G}{c}J^{ij},\label{03}\\
\epsilon^{\langle ikl}\partial^k{\cal E}^{lj\rangle}+\frac{1}{c}\frac{\partial{\cal B}^{ij}}{\partial t}&=&0,\label{04}
\eea
where $G$ is the gravitational constant, $\epsilon^{ikl}$ is the Levi-Civita symbol, $\rho^j$ is the vector mass density, $J^{ij}$ is the mass current density and $c$ is the speed of light. The quantities ${\cal E}^{ij}$, ${\cal B}^{ij}$ and $J^{ij}$ are the gravitoelectric field, the gravitomagnetic field and the mass current density, respectively. The symbol $\langle\cdots\rangle$ denotes symmetrization of the first and last indices i.e. $i$ and $j$.


A symmetric rank-2 tensor field, $\tilde{\cal A}$, with components ${\cal A}^{ij}$ is written such that the field ${\cal B}$ is expressed as
\bea
{\cal B}=\textrm{curl}\,\tilde{\cal A}.\label{B}
\eea
Using $\textrm{div}\, \textrm{curl}\,\tilde{\cal A}=\frac{1}{2}\mathrm{curl}\,\mathrm{div}\,\tilde{\cal A}$ where $\tilde{\cal A}$ is such that $\mathrm{div}\,\tilde{\cal A}=0$, then eq. (\ref{02}) is satisfied. It is possible to rewrite eq. (\ref{04}) as
\bea
\mathrm{curl}\left({\cal E}+\frac{1}{c}\frac{\partial\tilde{\cal A}}{\partial t}\right)=0,
\eea
where $\mathrm{curl}\,{\cal E}=\epsilon^{\langle ikl}\partial^k{\cal E}^{lj\rangle}$. Defining $\varphi$ as the GEM counterpart of the electromagnetic (EM) scalar potential $\phi$, the gravitoelectric field is written as
\bea
{\cal E}+\frac{1}{c}\frac{\partial\tilde{\cal A}}{\partial t}=-\mathrm{grad}\,\varphi.\label{E}
\eea

Then GEM fields are rewritten as
\bea
{\cal F}^{0ij}&=&{\cal E}^{ij},\\
{\cal F}^{ijk}&=&\epsilon^{ijl}{\cal B}^{lk},
\eea
where ${\cal F}^{\mu\nu\alpha}$ is the gravitoelectromagnetic tensor defined as 
\bea
{\cal F}^{\mu\nu\alpha}=\partial^\mu{\cal A}^{\nu\alpha}-\partial^\nu{\cal A}^{\mu\alpha},
\eea
with $\mu, \nu,\alpha=0, 1, 2, 3$. 

The dual GEM tensor is defined as
\bea
{\cal G}^{\mu\nu\alpha}=\frac{1}{2}\epsilon^{\mu\nu\gamma\sigma}\eta^{\alpha\rho}{\cal F}_{\gamma\sigma\rho},
\eea
where $\eta_{\mu\nu}=(+,-,-,-)$. 

Thus the GEM field equations are 
\bea
\partial_\mu{\cal F}^{\mu\nu\alpha}&=&\frac{4\pi G}{c}{\cal J}^{\nu\alpha},\\
\partial_\mu{\cal G}^{\mu\langle\nu\alpha\rangle}&=&0,
\eea
where ${\cal J}^{\nu\alpha}$ depends on quantities $\rho^i$ and $J^{ij}$ that are the mass and the current density, respectively. Then the GEM lagrangian density is 
\bea
{\cal L}_G=-\frac{1}{16\pi}{\cal F}_{\mu\nu\alpha}{\cal F}^{\mu\nu\alpha}-\frac{G}{c}\,{\cal J}^{\nu\alpha}{\cal A}_{\nu\alpha}.
\eea

Hereafter, we consider the lagrangian density for the free GEM field given by
\bea
{\cal L}_G=-\frac{1}{16\pi}{\cal F}_{\rho\sigma\theta}{\cal F}^{\rho\sigma\theta}. \label{LGEM}
\eea
Using this lagrangian density, the energy-momentum tensor is
\bea
\mathbb{T}^{\mu\nu}&=&\frac{\partial {\cal L}_G}{\partial(\partial_\mu A_{\lambda\alpha})}\partial^\nu A_{\lambda\alpha}-g^{\mu\nu}{\cal L}_G,\nonumber\\
&=&-\frac{1}{4\pi}{\cal F}^{\mu\lambda\alpha}\partial^\nu A_{\lambda\alpha}+\frac{1}{16\pi}g^{\mu\nu}{\cal F}_{\rho\sigma\theta}{\cal F}^{\rho\sigma\theta}.
\eea
Observe that this tensor is not symmetric. To symmetrize it we use ${\cal F}^{\mu\lambda\alpha}\partial^\nu A_{\lambda\alpha}=g^{\mu\rho}{\cal F}_{\rho\lambda\alpha}\partial^\nu A^{\lambda\alpha}$ and $\partial^\nu A^{\lambda\alpha}={\cal F}^{\nu\lambda\alpha}+\partial^\lambda A^{\nu\alpha}$, then
\bea
\mathbb{T}^{\mu\nu}=-\frac{1}{4\pi}g^{\mu\rho}{\cal F}_{\rho\lambda\alpha}{\cal F}^{\nu\lambda\alpha}-\frac{1}{4\pi}g^{\mu\rho}{\cal F}_{\rho\lambda\alpha}\partial^\lambda A^{\nu\alpha}+\frac{1}{16\pi}g^{\mu\nu}{\cal F}_{\rho\sigma\theta}{\cal F}^{\rho\sigma\theta}.
\eea
Using the definition
\bea
T^{\mu\nu}=\mathbb{T}^{\mu\nu}+\frac{1}{4\pi}g^{\mu\rho}{\cal F}_{\rho\lambda\alpha}\partial^\lambda A^{\nu\alpha},
\eea
the symmetric energy-momentum tensor for the GEM field is
\bea
T^{\mu\nu}=\frac{1}{4\pi}\left[-{\cal F}^\mu_{\lambda\alpha}{\cal F}^{\nu\lambda\alpha}+\frac{1}{4}g^{\mu\nu}{\cal F}_{\rho\sigma\theta}{\cal F}^{\rho\sigma\theta}\right].\label{EMT}
\eea

The canonical conjugated momentum related to $A_{\kappa\lambda}$ is
\bea
\pi^{\kappa\lambda}=\frac{\partial {\cal L}_G}{\partial(\partial_0 A_{\kappa\lambda})}=-\frac{1}{4\pi}{\cal F}^{0\kappa\lambda}.
\eea
Adopting the Coulomb gauge, where $A^{0i}=0$ and $\mathrm{div}\tilde{A}=\partial_i A^{ij}=0$, the covariant quantization is carried out and the commutation relation is
\bea
\left[A^{ij}({\bf x},t),\pi^{kl}({\bf x}',t)\right]&=&\frac{i}{2}\Bigl[\delta^{ik}\delta^{jl}-\delta^{il}\delta^{jk}-\frac{1}{\nabla^2}\Bigl(\delta^{jl}\partial^i\partial^k-\nonumber\\
&-&\delta^{jk}\partial^i\partial^l-\delta^{il}\partial^j\partial^k+\delta^{ik}\partial^j\partial^l\Bigl)\Bigl]\delta^3({\bf x}-{\bf x}').\label{CR}
\eea
Other commutation relations are zero.

Calculating the expectation value of $T^{\mu\nu}(x)$, given by eq. (\ref{EMT}), in the vacuum state is not possible due to the product of field operators at the same point of the space-time. To solve this problem we write
\bea
T^{\mu\nu}(x)&=&\frac{1}{4\pi}\lim_{x\rightarrow x'}\left\{\tau\left[-{\cal F}^\mu_{\lambda\alpha}(x){\cal F}^{\nu\lambda\alpha}(x')+\frac{1}{4}g^{\mu\nu}{\cal F}_{\rho\sigma\theta}(x){\cal F}^{\rho\sigma\theta}(x')\right]\right\},\nonumber\\
&=&\frac{1}{4\pi}\lim_{x\rightarrow x'}\left[-\mathbb{F}^\mu\,_{\lambda\alpha,}\,^{\nu\lambda\alpha}(x,x')+\frac{1}{4}g^{\mu\nu}\mathbb{F}^{\rho\sigma\theta,}\,_{\rho\sigma\theta}(x,x')\right],
\eea
where $\tau$ is the time order operator and
\bea
\mathbb{F}^{\alpha\kappa\gamma,\mu\nu\rho}(x,x')=\tau\left[{\cal F}^{\alpha\kappa\gamma}(x){\cal F}^{\mu\nu\rho}(x')\right].
\eea
Using the $\tau$ operator explicity
\bea
\mathbb{F}^{\alpha\kappa\gamma,\mu\nu\rho}(x,x')={\cal F}^{\alpha\kappa\gamma}(x){\cal F}^{\mu\nu\rho}(x')\theta(x_0-x_0')+{\cal F}^{\mu\nu\rho}(x'){\cal F}^{\alpha\kappa\gamma}(x)\theta(x_0'-x_0),
\eea
with $\theta(x_0-x_0')$ being the step function. In calculations that follow, we use the commutation relation, eq. (\ref{CR}), and 
\bea
\partial^\mu\theta(x_0-x_0')=n^\mu_0\delta(x_0-x_0'),
\eea
where $n^\mu_0=(1,0,0,0)$ is a time-like vector. Then we get
\bea
\mathbb{F}^{\alpha\kappa\gamma,\mu\nu\rho}(x,x')&=&\Gamma^{\alpha\kappa\gamma,\mu\nu\rho,\lambda\epsilon\omega\upsilon}(x,x')\tau\left[A_{\lambda\epsilon}(x)A_{\omega\upsilon}(x')\right]\nonumber\\
&+&I^{\kappa\gamma, \mu\nu\rho}(x,x')n^\alpha_0 \delta(x_0-x_0')-I^{\alpha\gamma, \mu\nu\rho}(x,x')n^\kappa_0 \delta(x_0-x_0'),
\eea
where 
\bea
\Gamma^{\alpha\kappa\gamma,\mu\nu\rho,\lambda\epsilon\omega\upsilon}(x,x')=\left(g^{\kappa\lambda}g^{\epsilon\gamma}\partial^\alpha-g^{\alpha\lambda}g^{\epsilon\gamma}\partial^\kappa\right)\left(g^{\nu\omega}g^{\rho\upsilon}\partial'^\mu-g^{\mu\omega}g^{\rho\upsilon}\partial'^\nu\right)
\eea
and
\bea
I^{\kappa\gamma, \mu\nu\rho}(x,x')&=&\left[A^{\kappa\gamma}(x), {\cal F}^{\mu\nu\rho}(x')\right]\nonumber\\
&=&\left[A^{\kappa\gamma}(x), \partial'^\mu A^{\nu\rho}(x')\right]-\left[A^{\kappa\gamma}(x), \partial'^\nu A^{\mu\rho}(x')\right]\nonumber\\
&=& \frac{i}{2}n^\mu_0\Bigl[\delta^{\kappa\nu}\delta^{\gamma\rho}-\delta^{\kappa\rho}\delta^{\gamma\nu}-\frac{1}{\nabla^2}\Bigl(\delta^{\gamma\rho}\partial^\kappa\partial^\nu-\delta^{\gamma\nu}\partial^\kappa\partial^\rho-\delta^{\kappa\rho}\partial^\gamma\partial^\nu+\nonumber\\
&+&\delta^{\kappa\nu}\partial^\gamma\partial^\rho\Bigl)\Bigl]\delta^3({\bf x}-{\bf x}') -\frac{i}{2}n^\nu_0\Bigl[\delta^{\kappa\mu}\delta^{\gamma\rho}-\delta^{\kappa\rho}\delta^{\gamma\mu}-\frac{1}{\nabla^2}\Bigl(\delta^{\gamma\rho}\partial^\kappa\partial^\mu-\nonumber\\
&-&\delta^{\gamma\mu}\partial^\kappa\partial^\rho-\delta^{\kappa\rho}\partial^\gamma\partial^\mu+\delta^{\kappa\mu}\partial^\gamma\partial^\rho\Bigl)\Bigl]\delta^3({\bf x}-{\bf x}').
\eea

Then the energy-momentum tensor is
\bea
T^{\mu\nu}(x)=-\frac{1}{4\pi}\lim_{x\rightarrow x'}\left\{\Delta^{\mu\nu,\lambda\epsilon\omega\upsilon}(x,x')\tau\left[A_{\lambda\epsilon}(x)A_{\omega\upsilon}(x')\right]\right\},
\eea
where
\bea
\Delta^{\mu\nu,\lambda\epsilon\omega\upsilon}(x,x')=\Gamma^\mu\,_{\rho\alpha,}\,^{\nu\rho\alpha,\lambda\epsilon\omega\upsilon}(x,x')-\frac{1}{4}g^{\mu\nu}\Gamma^{\rho\sigma\theta,}\,_{\rho\sigma\theta,}\,^{\lambda\epsilon\omega\upsilon}(x,x').
\eea

The vacuum expectation value of the energy-momentum tensor is
\bea
\langle T^{\mu\nu}(x)\rangle &=& \langle 0|T^{\mu\nu}(x)|0\rangle,\nonumber\\
&=&-\frac{1}{4\pi}\lim_{x\rightarrow x'}\left\{\Delta^{\mu\nu,\lambda\epsilon\omega\upsilon}(x,x')\left\langle 0\left| \tau\left[A_{\lambda\epsilon}(x)A_{\omega\upsilon}(x')\right]\right| 0\right\rangle\right\}.
\eea
Using the graviton propagator given by
\bea
D_{\lambda\epsilon\omega\upsilon}(x-x')&=&\left\langle 0\left| \tau\left[A_{\lambda\epsilon}(x)A_{\omega\upsilon}(x')\right]\right| 0\right\rangle,\nonumber\\
&=&i\,\Theta_{\lambda\epsilon\omega\upsilon}G_0(x-x'),
\eea
where
\bea
\Theta_{\lambda\epsilon\omega\upsilon}=\frac{1}{2}\left(g_{\lambda\omega}g_{\epsilon\upsilon}+g_{\lambda\upsilon}g_{\epsilon\omega}-g_{\lambda\epsilon}g_{\omega\upsilon}\right),
\eea
and $G_0(x-x')$ is the massless scalar field propagator given in eq. (\ref{G0}). Then the vacuum expectation value of $T^{\mu\nu}(x)$ becomes
\bea
\langle T^{\mu\nu}(x)\rangle=-\frac{i}{8\pi}\lim_{x\rightarrow x'}\left\{\Gamma^{\mu\nu}(x,x')G_0(x-x')\right\},
\eea
with
\bea
\Gamma^{\mu\nu}(x,x')=8\left(\partial^\mu\partial'^\nu-\frac{1}{4}g^{\mu\nu}\partial^\rho\partial'_\rho\right).\label{Gamma}
\eea

Following the tilde conjugation rules, the vacuum average of the energy-momentum tensor in terms of the $\alpha$-dependent fields is
\bea
\langle T^{\mu\nu(ab)}(x;\alpha)\rangle=-\frac{i}{8\pi}\lim_{x\rightarrow x'}\left\{\Gamma^{\mu\nu}(x,x')G_0^{(ab)}(x-x';\alpha)\right\}.
\eea

The physical energy-momentum tensor is given by
\bea
{\cal T}^{\mu\nu (ab)}(x;\alpha)=\langle T^{\mu\nu(ab)}(x;\alpha)\rangle-\langle T^{\mu\nu(ab)}(x)\rangle.
\eea
With this procedure a measurable physical quantity is obtained. Explicitly
\bea
{\cal T}^{\mu\nu (ab)}(x;\alpha)=-\frac{i}{8\pi}\lim_{x\rightarrow x'}\left\{\Gamma^{\mu\nu}(x,x')\overline{G}_0^{(ab)}(x-x';\alpha)\right\},\label{VEV}
\eea
where 
\bea
\overline{G}_0^{(ab)}(x-x';\alpha)=G_0^{(ab)}(x-x';\alpha)-G_0^{(ab)}(x-x').
\eea
In the Fourier representation 
\bea
\overline{G}_0^{(ab)}(x-x';\alpha)=\int \frac{d^4k}{(2\pi)^4}e^{-ik(x-x')}\overline{G}_0^{(ab)}(k;\alpha),
\eea
such that the important component of $\overline{G}_0^{(ab)}(k;\alpha)$ is
\bea
\overline{G}_0^{(11)}(k;\alpha)=v^2(\alpha)\left[G_0(k)-G_0^*(k)\right].
\eea

The generalized Bogoliubov transformation \cite{GBT} is
\bea
v^2(k_\alpha;\alpha)=\sum_{s=1}^d\sum_{\lbrace\sigma_s\rbrace}2^{s-1}\sum_{l_{\sigma_1},...,l_{\sigma_s}=1}^\infty(-\eta)^{s+\sum_{r=1}^sl_{\sigma_r}}\,\exp\left[{-\sum_{j=1}^s\alpha_{\sigma_j} l_{\sigma_j} k^{\sigma_j}}\right],\label{BT}
\eea
where $d$ is the number of compactified dimensions, $\eta=1(-1)$ for fermions (bosons) and $\lbrace\sigma_s\rbrace$ denotes the set of all combinations with $s$ elements. In order to obtain physical conditions at finite temperature and spatial confinement, $\alpha_0$ has to be taken as a positive real number, while $\alpha_n$
for $n=1,2,\cdots, d-1$ must be pure imaginary of the form $iL_n$.

As a first application, consider the thermal effect that appears for $\alpha=(\beta,0,0,0)$, such that
\bea
v^2(\beta)=\sum_{j_0=1}^\infty e^{-\beta k^0 j_0}.
\eea
Thus
\bea
\overline{G}_0^{(11)}(x-x';\alpha)&=&\int \frac{d^4k}{(2\pi)^4}e^{-ik(x-x')}\sum_{j_0=1}^\infty e^{-\beta k^0 j_0}\left[G_0(k)-G_0^*(k)\right],\nonumber\\
&=&2\sum_{j_0=1}^\infty G_0\left(x-x'-i\beta j_0 n_0\right),\label{1GF}
\eea
where $n_0^\mu=(1,0,0,0)$. Then the vacuum expectation value of the energy-momentum tensor at finite temperature, eq. (\ref{VEV}), becomes
\bea
{\cal T}^{\mu\nu(11)}(x;\beta)&=&-\frac{2i}{\pi}\lim_{x\rightarrow x'}\left\lbrace \sum_{j_0=1}^\infty\left(\partial^\mu\partial'^\nu-\frac{1}{4}g^{\mu\nu}\partial^\rho\partial'_\rho\right) G_0\left(x-x'-i\beta j_0 n_0\right) \right\rbrace,\nonumber\\
&=&\frac{1}{\pi^3}\sum_{j_0=1}^\infty\frac{1}{\beta^4j_0^4}\left(4n^\mu_0 n^\nu_0-g^{\mu\nu}\right),\label{SB}
\eea
where eq. (\ref{Gamma}) has been used. Using the Riemann Zeta function
\bea
\zeta(4)=\sum_{j_0=1}^\infty\frac{1}{j_0^4}=\frac{\pi^4}{90},\label{zetaf}
\eea
eq. (\ref{SB}) becomes
\bea
{\cal T}^{\mu\nu(11)}(x;\beta)=\frac{\pi}{90\beta^4}\left(4n^\mu_0 n^\nu_0-g^{\mu\nu}\right).
\eea
This leads to the Stefan-Boltzmann law for the GEM field
\bea
E(T)={\cal T}^{00(11)}(x;\beta)=\frac{\pi}{30} T^4.
\eea

\section{Casimir effect for GEM field}

Now the Casimir effect at zero and finite temperature is calculated. With different choices for the $\alpha$ parameter the energy-momentum tensor is calculated.

\subsection{Casimir effect at zero temperature}

For $\alpha=(0,0,0,iL)$, the Bogoliubov transformation is
\bea
v^2(L)=\sum_{l_3=1}^\infty e^{-iLk^3l_3}.
\eea
The Green function is
\bea
\overline{G}_0^{(11)}(x-x';L)&=&\int \frac{d^4k}{(2\pi)^4}e^{-ik(x-x')}\sum_{l_3=1}^\infty e^{-iLk^3 l_3}\left[G_0(k)-G_0^*(k)\right],\nonumber\\
&=&2\sum_{l_3=1}^\infty G_0\left(x-x'-Ll_3z\right).\label{2GF}
\eea
A sum over $l_3$, for $L=2d$, defines the nontrivial part of the Green function with the Dirichlet boundary condition. Then the energy-momentum tensor becomes
\bea
{\cal T}^{\mu\nu(11)}(x;d)&=&-\frac{2i}{\pi}\lim_{x\rightarrow x'}\left\lbrace \sum_{l_3=1}^\infty\left(\partial^\mu\partial'^\nu-\frac{1}{4}g^{\mu\nu}\partial^\rho\partial'_\rho\right) G_0\left(x-x'-2dl_3z\right) \right\rbrace,\nonumber\\
&=&-\frac{1}{16\pi^3}\sum_{l_3=1}^\infty\frac{1}{d^4l_3^4}\left(g^{\mu\nu}+4n^\mu_3 n^\nu_3\right).\label{zerotemp}
\eea
Using eq. (\ref{zetaf}) we have
\bea
{\cal T}^{\mu\nu(11)}(x;d)=-\frac{\pi}{1440d^4}\left(g^{\mu\nu}+4n^\mu_3 n^\nu_3\right),
\eea
where $n^\mu_3=(0,0,0,1)$. The Casimir energy and pressure for the GEM field are
\bea
E(d)&=&{\cal T}^{00(11)}(x;d)=-\frac{\pi}{1440d^4},\nonumber\\
P(d)&=&{\cal T}^{33(11)}(x;d)=-\frac{\pi}{480d^4}.
\eea
The negative sign shows that the Casimir force between the plates is attractive, similar to the case of the electromagnetic field.

\subsection{Casimir effect at finite temperature}

For parallel plates with $L=2d$, the effect of temperature is introduced by taking $\alpha=(\beta, 0, 0,i2d)$. In this case the Bogoliubov transformation, eq. (\ref{BT}), becomes
\bea
v^2(k^0,k^3;\beta,d)&=&v^2(k^0;\beta)+v^2(k^3;d)+2v^2(k^0;\beta)v^2(k^3;d),\nonumber\\
&=&\sum_{j_0=1}^\infty e^{-\beta k^0j_0}+\sum_{l_3=1}^\infty e^{-iLk^3l_3}+2\sum_{j_0,l_3=1}^\infty e^{-\beta k^0j_0-iLk^3l_3}.
\eea
The Green function, corresponding to the first two terms, is given in eq. (\ref{1GF}) and in eq. (\ref{2GF}), respectively. The Green function associated with the third term is
\bea
\overline{G}_0^{(11)}(x-x';\beta,d)&=&2\int \frac{d^4k}{(2\pi)^4}e^{-ik(x-x')}\sum_{j_0,l_3=1}^\infty e^{-\beta k^0j_0-iLk^3l_3}\left[G_0(k)-G_0^*(k)\right],\nonumber\\
&=&4\sum_{j_0,l_3=1}^\infty G_0\left(x-x'-i\beta j_0n-2dl_3z\right).\label{3GF}
\eea
Then the energy-momentum tensor is given as
\bea
{\cal T}^{\mu\nu(11)}(\beta;d)&=&-\frac{1}{\pi^3}\Biggl\lbrace\sum_{j_0=1}^\infty\frac{g^{\mu\nu}-4n^\mu_0n^\nu_0}{(\beta j_0)^4}+\sum_{l_3=1}^\infty\frac{g^{\mu\nu}+4n^\mu_3n^\nu_3}{(2dl_3)^4}\nonumber\\
&+&2\sum_{j_0,l_3=1}^\infty\frac{(\beta j_0)^2[g^{\mu\nu}-4n^\mu_0n^\nu_0]+(2dl_3)^2[g^{\mu\nu}+4n^\mu_3n^\nu_3]}{[(\beta j_0)^2+(2dl_3)^2]^3}\Biggl\rbrace,\label{Combined}
\eea
and the Casimir energy and pressure are given, respectively, by
\bea
E(\beta, d)={\cal T}^{00(11)}(\beta;d)&=&\frac{\pi}{30\beta^4}-\frac{\pi}{1440d^4}+\frac{2}{\pi^3}\sum_{j_0,l_3=1}^\infty\frac{3(\beta j_0)^2-(2dl_3)^2}{[(\beta j_0)^2+(2dl_3)^2]^3},\label{ED}\\
P(\beta, d)={\cal T}^{33(11)}(\beta;d)&=&\frac{\pi}{90\beta^4}-\frac{\pi}{480d^4}+\frac{2}{\pi^3}\sum_{j_0,l_3=1}^\infty\frac{(\beta j_0)^2-3(2dl_3)^2}{[(\beta j_0)^2+(2dl_3)^2]^3}.\label{P}
\eea
The first term is the Stefan-Boltzmann law. The second and last term are Casimir effect at zero and finite temperature, respectively.

In order to get the physical quantities involved in the last term of the eq. (\ref{Combined}), this term is written as
\bea
{\cal T}^{\mu\nu(11)}_c(\beta;d)&=&-\frac{2}{\pi^3}\sum_{j_0,l_3=1}^\infty\frac{(\beta j_0)^2[g^{\mu\nu}-4n^\mu_0n^\nu_0]+(2dl_3)^2[g^{\mu\nu}+4n^\mu_3n^\nu_3]}{[(\beta j_0)^2+(2dl_3)^2]^3},\nonumber\\
&=&-\frac{2}{\pi^3}\sum_{j_0,l_3=1}^\infty\Bigl\lbrace \frac{g^{\mu\nu}}{[(\beta j_0)^2+(2dl_3)^2]^2}-4\frac{[(\beta j_0)^2n^\mu_0 n^\nu_0-(2dl_3)^2n^\mu_3 n^\nu_3]}{[(\beta j_0)^2+(2dl_3)^2]^3} \Bigl\rbrace,
\eea
Following \cite{Brown}, for $\xi=\frac{d}{\beta}$ we define ,
\bea
f(\xi)=-\frac{1}{8\pi^3}\sum_{j_0,l_3=1}^\infty\frac{(2\xi)^4}{[j_0^2+(2l_3\xi)^2]^2},
\eea
and
\bea
S(\xi)=-\frac{df(\xi)}{d\xi}=\frac{2^3}{\pi^3}\sum_{j_0,l_3=1}^\infty\frac{\xi^3j_0^2}{[j_0^2+(2l_3\xi)^2]^3}.
\eea
Then the energy-momentum tensor becomes
\bea
{\cal T}^{\mu\nu(11)}_c(\beta;d)=\frac{1}{d^4}f(\xi)\left[g^{\mu\nu}+4n^\mu_3n^\nu_3\right]+\frac{1}{\beta d^3}\left[n^\mu_0n^\nu_0+n^\mu_3n^\nu_3\right]S(\xi).
\eea
The energy density is given by
\bea
E_c(\beta,d)=\frac{1}{d^4}\left[f(\xi)+\xi S(\xi)\right].
\eea
Here, functions $f(\xi)$ and $S(\xi)$, describe the free energy density and entropy density, respectively, for gravitons .

\section{Conclusions}

The Casimir effect for electromagnetic (EM) field between metal plates has been measured at zero temperature. In addition this effect at finite temperature has been established when the separation between plates is of the order of micrometers. Is it interesting to calculate the Casimir effect for gravitational fields? Some linearized models of gravitation fields have been utilized to estimate the Casimir effect. Here we use the gravitoelectromagnetism field.

Gravitoelectromagnetism (GEM) is based on an analogy with electromagnetism. Here we use the version where the Weyl tensor is divided into electric ${\cal E}_{ij}=-C_{0i0j}$ and magnetic ${\cal B}_{ij}=\frac{1}{2}\epsilon_{ikl}C^{kl}_{0j}$ components. Finite temperature is introduced using Thermo Field Dynamics (TFD), a real-time finite temperature field theory. Using this formalism the energy-momentum tensor is obtained at finite temperature. Then the Casimir energy and pressure are calculated. The role of temperature is small. These results are similar to the case of electromagnetic field. 

It is important to point out that although these results are similar there are important difference between two theories. For example, electromagnetic fields are vectors whereas GEM fields are tensors. The electromagnetic field is opaque to the plates and then its contribution for the Casimir effect is measurable. The gravitational field for conventional plates is small. However an experimental proposal \cite{Quach} to measure the gravitational Casimir effect using GEM field is developed. Our results indicate that the combined effect of temperature and spatial compactification give a new constraint for the gravitational Casimir effect.

\section*{Acknowledgments}

This work by A. F. S. is supported by CNPq projects 476166/2013-6 and 201273/2015-2. We thank Physics Department, University of Victoria for access to facilities.


\begin{thebibliography}{99}
\bibitem{Casimir} H. G. B. Casimir, Proc. K. Ned. Akad. Wet. {\bf 51}, 793 (1948).
\bibitem{Milton} K. A. Milton, The Casimir Effect, Physical Manifestations of Zero-Point Energy, World Scientific, Singapore, (2001).
\bibitem{Milonni} P. W. Milonni, The Quantum Vaccum: An Introduction to Quantum Electrodynamics, Academic Press, New York, (1994).
\bibitem{Plunien} G. Plunien, B. Muller and W. Greiner, Phys. Rep. {\bf 134}, 89 (1986).
\bibitem{Bordag} M. Bordag, U. Mohideen and V. M. Mostepanenko, Phys. Rep. {\bf 353}, 1 (2001).
\bibitem{Sparnaay} M. J. Sparnaay, Physica {\bf 24}, 751 (1958).
\bibitem{Lamoreaux} S. K. Lamoreaux, Phys. Rev. Lett. {\bf 28}, 5 (1997).
\bibitem{Mohideen} U. Mohideen and A. Roy, Phys. Rev. Lett. {\bf 81}, 21 (1998).
\bibitem{Lamoreaux1}S. K. Lamoreaux, Am. J. Phys. {\bf 67}, 850 (1999).
\bibitem{Bordag1} M. Bordag, J. Phys. A {\bf 39}, 6173 (2006).
\bibitem{Orlando}M. T. D. Orlando et al., J. Phys. A: Math. Theor. {\bf 42}, 025502 (2009).
\bibitem{Mehra} J. Mehra, Physica (Amsterdam) {\bf 37}, 145 (1967).
\bibitem{Khanna0} J. C. da Silva, F. C. Khanna, A. Matos Neto and A. E. Santana, Phys. Rev. A {\bf 66}, 052101 (2002).
\bibitem{Ademir} H. Belich, L. M. Silva, J. A. Helayel-Neto and A. E. Santana,  Phys. Rev. D {\bf 84}, 045007 (2011). 
\bibitem{Umezawa1}Y. Takahashi and H. Umezawa, Coll. Phenomena {\bf 2}, 55 (1975); Int. Jour. Mod. Phys. B {\bf 10}, 1755 (1996).
\bibitem{Umezawa2}Y. Takahashi, H. Umezawa and H. Matsumoto, Thermofield Dynamics and Condensed States, North-Holland, Amsterdan, (1982); F. C. Khanna, A. P. C. Malbouisson, J. M. C. Malboiusson and A. E. Santana, Themal quantum field theory: Algebraic aspects and applications, World Scientific, Singapore, (2009).
\bibitem{Umezawa22} H. Umezawa, Advanced Field Theory: Micro, Macro and Thermal Physics, AIP, New York, (1993).
\bibitem{Khanna1} A. E. Santana and F. C. Khanna, Phys. Lett. A {\bf 203}, 68 (1995).
\bibitem{Khanna2} A. E. Santana, F. C. Khanna, H. Chu, and C. Chang, Ann. Phys. {\bf 249}, 481 (1996).
\bibitem{Fulling} S. A. Fulling, K. A. Milton, P. Parashar, A. Romeo, K. V. Shajesh, and J. Wagner, Phys. Rev. D {\bf 76}, 025004 (2007).
\bibitem{Parashar} K. A. Milton, P. Parashar, K. V. Shajesh and J. Wagner, J. Phys. A: Math. Theor. {\bf 40}, 10935 (2007).
\bibitem{Zhuk} A. Zhuk and H. Kleinert, Teor. Mat. Fiz., {\bf 109}, 307 (1996).
\bibitem{Bezerra1} V. B. Bezerra, V. M. Mostepanenko, H. F. Mota, and C. Romero, Phys. Rev. D {\bf 84}, 104025 (2011).
\bibitem{Bezerra2} V. B. Bezerra, H. F. Mota, and C. R. Muniz, Phys. Rev. D {\bf 89}, 024015 (2014).
\bibitem{Bezerra3} H. F. Mota and V. B. Bezerra, Phys. Rev. D {\bf 92}, 124039 (2015).
\bibitem{Bezerra4} C. R. Muniz, V. B. Bezerra and M. S. Cunha, Ann. Phys. {\bf 359}, 55 (2015).
\bibitem{Bezerra5} C. R. Muniz, V. B. Bezerra and M. S. Cunha, Phys. Rev. D {\bf 88}, 104035 (2013).
\bibitem{Quach} J. Q. Quach, Phys. Rev. Lett. {\bf 114}, 081104 (2015).
\bibitem{Maxwell}J. C. Maxwell, Phil. Trans. Soc. Lond. {\bf 155}, 492 (1865).
\bibitem{Heaviside1} O. Heaviside, Electrician {\bf 31}, 259 (1893).
\bibitem{Heaviside2} O. Heaviside, Electrician {\bf 31}, 281 (1893).
\bibitem{Wheeler}I. Ciufolini and J. A. Wheeler, Gravitation and Inertia, Princeton University Press, Princeton, (1951).
\bibitem{Nordtvedt}K. Nordtvedt, Rev. Phys. Lett. {\bf 61}, 2647 (1988).
\bibitem{Soffel}M. Soffel, S. Klioner, J. Muller and L. Biskupek, Phys. Rev. D {\bf 78}, 024033 (2008).
\bibitem{Everitt}C. W. F. Everitt et al., Phys. Rev. Lett. {\bf 106}, 221101 (2011).
\bibitem{Khanna}J. Ramos, M. de Montigny and F. C. Khanna, Gen. Rel. Grav. {\bf 42}, 2403 (2010).
\bibitem{Santana1}A. E. Santana, A. Matos Neto, J. D. M. Vianna and F.C. Khanna, Physica A {\bf 280}, 405 (2000).
\bibitem{Santana2}F. C. Khanna, A. P. C Malbouisson, J. M. C. Malbouisson and A. E. Santana, Ann. Phys. {\bf 324}, 1931 (2009).
\bibitem{GBT}F. C. Khanna, A. P. C Malbouisson, J. M. C. Malbouisson and A. E. Santana, Ann. Phys. {\bf 326}, 2634 (2011). 
\bibitem{Brown} L. S. Brown and G. J. Maclay, Phys. Rev. {\bf 184}, 1272 (1969).
\end{thebibliography}
\end{document}